\documentclass[prl,aps,twocolumn,showpacs,epsf]{revtex4}
\usepackage{graphicx}
\begin{document}

\title{Stripe formation driven by space noncommutativity in quantum Hall systems}
\author{Guo-Zhu Liu and Sen Hu \\
{\small {\it Department of Mathematics, University of Science and
Technology of China, Hefei, Anhui, 230026, P. R. China }}}

\begin{abstract}
We propose that the transport anisotropy observed in half-filled
high Landau levels ($N \geq 2$) is caused by the space
noncommutativity effect, namely the Heisenberg uncertainty
relation between the spatial coordinates of electrons. The stripe
corresponds to a limit that one coordinate of a large number of
particles is fixed at a certain value while its conjugate
coordinate is completely uncertain. We make a renormalization
group analysis and find that the noncommutativity effect is able
to drive the stripe formation only at half-fillings $\nu = 9/2,
11/2, 13/2,$ etc., in agreement with experiments.
\end{abstract}

\pacs{71.70.Di, 73.43.-f, 11.10.Hi}

\maketitle


The two-dimensional electron gas in a perpendicular magnetic field
exhibits a class of fascinating phenomena at half Landau level
(LL) fillings. At half-fillings, an electron captures two flux
tubes to form a composite fermion \cite{Jain, Heinonen}, which
feels effectively zero net magnetic field. At $\nu = 1/2$ and
$3/2$ of $N = 0$ LL, the composite fermions constitute a gapless
Fermi liquid \cite{Halperin, Willett1}. At $\nu = 5/2$ and $7/2$
of $N = 1$ LL, the composite fermions are expected to undergo a
Cooper pair instability and condense into an ordered
superconducting state. Thus the fractional quantum Hall effect
state observed in experiments \cite{Willett2} can be regarded as a
$p$-wave BCS superconductor \cite{Moore, Greiter, Morf, Scarola}.
The Meissner effect of superconductor accounts for the
incompressibility. The Cooper pairing picture received strong
supports from extensive numerical calculations \cite{Morf,
Scarola}. At half-fillings $\nu = 9/2, 11/2, 13/2$ etc. of higher
LLs ($N \geq 2$), a strong transport anisotropy was observed
\cite{Lilly, Du99}. Specifically, the longitudinal resistance
shows a high peak in one current direction and a deep minimum in
its orthogonal direction, indicating that the electrons organize
themselves into stripes for some reason. The success of composite
fermion theory in understanding the half-fillings of $N = 0$ and
$N = 1$ LLs motivates us to generalize it to the half-fillings of
higher LLs ($N \geq 2$). The question is: if composite fermions do
exist at $9/2, 11/2, 13/2,$ etc., what is the underlying mechanism
that drives the stripe formation? We propose that the space
noncommutativity is the best candidate.

It is known that a Heisenberg uncertainty relation exists between
the spatial coordinates \cite{Girvin, Bellissard}. This can be
seen from the electron wave function
\begin{equation}
\psi_{nk}=L^{-1/2}e^{iky}H_{n}(x+k\ell^{2}) \exp
(-\frac{1}{2\ell^{2}} \left(x+k\ell^{2}\right)^{2}),
\end{equation}
where $H_{n}$ is the $n$th Hermite polynomial, $\ell = \sqrt{\hbar
c/eB}$ is the magnetic length and $L$ is the sample size. The
guiding center along the $x$-axis is determined by $k$, the
$y$-component of momentum. Since $y$ does not commute with the
$y$-momentum, $x$ and $y$ do not commute and $[x,y] = i \theta$,
with a noncommutative parameter $\theta$ that is of dimension
(length)$^{2}$. The idea of space-time noncommutativity or
space-time quantization \cite{Snyder} was introduced into physics
about sixty years ago. Its physical implications \cite{Douglas}
and mathematical structures \cite{Connes94} have been studied
extensively, especially in the past several years. However, up to
date it is not clear whether we live in this kind of world or not.
The search for unambiguous experimental evidence of
noncommutativity in high-energy physics appears to be quite
difficult due to the ultra-high energy scale. On the contrary, it
should be easier to detect its effects in condensed matter
physics.

The Heisenberg uncertainty principle requires that $x$ and $y$ can
only be determined simultaneously to the extend restricted by
$\theta$. If one squeezes $\Delta x \rightarrow 0$, then $\Delta
y$ must tend to $\infty$, and {\em vice} {\em versa}. For a single
particle, if its $x$-coordinate is fixed, then its $y$-coordinate
is completely arbitrary. For a many-body system, if a
macroscopically large number of particles have the same
$x$-coordinate, then their motions are all confined to one
straight line along the $y$-axis. This state, when thermally
stable, can be identified as a stripe state. From the wave
function (1), we know that if the fermions in a given LL have the
same $k$, then their guiding centers would have the same
$x$-coordinate. When stripes come into being, the fermions in one
stripe would have the same wave functions. This is in
contradiction to the Pauli exclusion principle. However, this
conflict dissolves automatically if the composite fermions form
Cooper pairs and behave effectively like bosons. It is reasonable
to assume that the composite fermions form Cooper pairs at filling
factors $\nu = n/2$ with odd integer $n \geq 5$. In case of
insignificant noncommutative effect, the Cooper pairs condense
into a uniform superconducting state at zero temperature. If the
noncommutative effect is strong, then a nonuniform or stripe state
is more favorable. We study the first order phase transition to
the stripe state using the renormalization group (RG) analysis
within a noncommutative Ginzburg-Landau (GL) model. We find that
the filling factor has a critical value, only beyond which the
system develops stripe solutions. The critical value $n_{c}/2$
lies between $7/2$ and $9/2$, in agreement with experiments.

We write down a noncommutative GL model
\begin{equation}
S = - \int d^{3}x \left[\left(\nabla
\phi\right)^{2}+m^{2}\phi^{\ast} \phi + \frac{1}{4}V(\phi^{\ast},
\phi) \right],
\end{equation}
with the potential
\begin{equation}
V(\phi^{\ast},\phi)=g_{1}\phi^{\ast} \star \phi \star \phi^{\ast}
\star \phi + g_{2}\phi^{\ast} \star \phi^{\ast} \star \phi \star
\phi.
\end{equation}
The complex scalar field $\phi (x)$ represents the Cooper pair of
composite fermions. The noncommutative relation is $\left[x^{\mu},
x^{\nu} \right]_{\star}=x^{\mu}\star x^{\nu} - x^{\nu}\star
x^{\mu} = i \Theta^{\mu\nu}$. Here, the noncommutative relation
exists only between spatial coordinates, then the only nonzero
components of the anti-symmetric matrix $\Theta^{\mu\nu}$ are
$\Theta^{12} = - \Theta^{21} = \theta$. The product of two
functions is defined by \cite{Douglas}
\begin{equation}
(f\star g)(x) =
e^{\frac{i}{2}\Theta_{\mu\nu}\partial_{x}^{\mu}\partial_{y}^{\nu}}f(x)g(y)|_{y=x}.
\end{equation}
Note that there are two possible noncommutative quartic
interaction terms with two coupling constants which should be
treated on the same footing.

The commutative GL model enters a uniform ordered state at low
temperatures. When the noncommutative effect becomes strong, a
stripe state is possible \cite{Sondhi, Wu}. A powerful tool to
study the phase transition to stripe state is the modern RG
analysis developed by Shankar \cite{Shankar} and independently by
Polchinski \cite{Polchinski}. The stripe phase is characterized by
the appearance of a $\sim k^{4}\phi^{\ast}\phi$ term which leads
to a multi-critical Lifshitz point \cite{Luban}. Such a term is
generated when the system develops an anomalous dimension $\gamma
\leq -1$ for the scalar field \cite{Wu}. Its basic idea is to
identify the fixed point of the action and then study the
interaction terms. The methods are standard and have been
presented previously \cite{Wu, Shankar, Polchinski}.

The free field action $S_{0}$ in momentum space is
\begin{equation}
S_{0} = - \int_{|k|<\Lambda} \frac{d^{3}k}{(2\pi)^{3}}
k^{2}\phi^{\ast} (-k) \phi (k).
\end{equation}
In modern RG theory, ultraviolet cutoff $\Lambda$ is interpreted
as the energy scale above which the physics has no effects on the
physics below it. We separate $\phi (k)$ into slow modes $\phi_{s}
(k) = \phi (k)$ for $|k| < \Lambda/s$ and fast modes $\phi_{f} (k)
= \phi (k)$ for $\Lambda/s \leq |k| \leq \Lambda$ with $s$ a
number greater than unity. The fast modes in the partition
function $Z=\int \mathcal{D} \phi e^{S}$ can be integrated out,
leaving
\begin{equation}
S^{\prime}_{0} = - \int_{|k|<\Lambda/s} \frac{d^{3}k}{(2\pi)^{3}}
k^{2} \phi^{\ast}_{s} (-k) \phi_{s} (k).
\end{equation}
The free action $S_{0}$ is a fixed point in the sense that it is
invariant under the RG transformations
\begin{eqnarray}
k^{\prime} = sk, \qquad \phi^{\prime} (k^{\prime}) =
s^{-\frac{5}{2}} \phi_{s} (k).
\end{eqnarray}

We next consider the mass term. After integrating out the fast
modes and making RG transformation, we have
\begin{equation}
S^{\prime}_{2} = - m^{2}s^{2} \int_{|k^{\prime}|<\Lambda}
\frac{d^{3}k^{\prime}}{(2\pi)^{3}} \phi^{\ast \prime}(-k)
\phi^{\prime}(k).
\end{equation}
Since $r^{\prime}\equiv m^{\prime 2}=s^{2}r\equiv s^{2}m^{2}$, the
mass term is enhanced under RG transformation and hence is
relevant.

The quartic interaction term is
\begin{equation}
S_{4} = - \frac{1}{4} \int_{|k| < \Lambda} \phi^{\ast}(1) \phi(2)
\phi^{\ast}(3) \phi(4)u(1234).
\end{equation}
Here, $\int_{|k|<\Lambda}$ denotes
$\int_{|k|<\Lambda}\prod^{4}_{i=1}\frac{d^{3}k^{\prime}_{i}}{(2\pi)^{3}}
\delta (\sum_{i}k^{\prime}_{i})$. The vertex function $u(1234)$
has the form
\begin{equation}
g_{1}\cos (\frac{k_{1}\wedge k_{2}}{2}+\frac{k_{3}\wedge
k_{4}}{2}) + g_{2}\cos (\frac{k_{1}\wedge k_{3}}{2}) \cos
(\frac{k_{2}\wedge k_{4}}{2}).
\end{equation}
To eliminate the fast modes, we adopt the standard cumulant
expansion method and write the action as
\begin{equation}
S^{\prime}_{4} = \langle S_{4}\rangle_{0f} +
\frac{1}{2}\left(\langle S_{4}^{2}\rangle_{0f} - \langle
S_{4}\rangle_{0f}^{2}\right) + \cdots,
\end{equation}
with $\langle \ldots \rangle_{0f}$ denoting functional integration
over the fast modes.

The leading term has the form
\begin{equation}
\langle \int (\phi_{s}+\phi_{f})_{1}^{\ast}(\phi_{s}+\phi_{f})_{2}
(\phi_{s}+\phi_{f})_{3}^{\ast} (\phi_{s}+\phi_{f})_{4} u(1234)
\rangle_{0f}.
\end{equation}
In all the sixteen terms, we only care about the term with all
slow modes and four terms with
$\phi^{\ast}_{s}\phi_{s}\phi^{\ast}_{f}\phi_{f}$. All other terms
either vanish or contribute only a constant.

The term with all slow modes is
\begin{equation}
S^{\prime}_{4,\mathrm{t}} = - \frac{1}{4} s
\int_{|k^{\prime}|<\Lambda} \phi^{\prime \ast}(1^{\prime})
\phi^{\prime}(2^{\prime}) \phi^{\prime \ast}(3^{\prime})
\phi^{\prime}(4^{\prime})u(1^{\prime}2^{\prime}3^{\prime}4^{\prime})
\end{equation}
after RG transformations. This is the tree-level quartic term and
is a relevant perturbation. The noncommutative star product
structure does not change under the RG transformations \cite{Wu}
and hence $g^{\prime}_{1}=sg_{1}$ and $g^{\prime}_{2}=sg_{2}$.

We next consider the four terms that contain two slow modes and
two fast modes. After calculating the loop integral of fast modes,
we have
\begin{eqnarray}
&& - \frac{2}{\pi^{2}} (g_{1}+g_{2}) \Lambda(1-\frac{1}{s})
\int_{|k|<\Lambda/s} \phi^{\ast}_{s}(-k)\phi_{s}(k) \nonumber \\
&& + \frac{g_{2}}{6\pi^{2}}\theta^{2}\Lambda^{3}(1-\frac{1}{s})
\int_{|k|<\Lambda/s} {\bf k}^{2}\phi^{\ast}_{s}(-k)\phi_{s}(k).
\end{eqnarray}
The first term only corrects the mass term, so we do not care
about it. The second term alters the kinetic term to
\begin{equation}
- \left[1-\frac{u_{2}}{6\pi^{2}}(\theta \Lambda^{2})^{2} t \right]
\int_{|k|<\Lambda/s} {\bf k}^{2} \phi^{\ast}_{s}(-k)\phi_{s}(k).
\end{equation}
Here we define a dimensionless parameter $u_{2}=g_{2}\Lambda^{-1}$
and write $s=1+t$ with $t$ infinitesimal. If we define
\begin{equation}
\gamma = -\frac{u_{2}}{6\pi^{2}}(\theta \Lambda^{2})^{2},
\end{equation}
then, using the formula $1+\gamma t \approx s^{\gamma}$, we have
\begin{equation}
- s^{\gamma} \int_{|k|<\Lambda/s} {\bf k}^{2}
\phi^{\ast}_{s}(-k)\phi_{s}(k).
\end{equation}
Now make the RG transformation $k^{\prime} = sk$, then we know
that the scalar field must transform as $\phi^{\prime}(k^{\prime})
= s^{-\frac{5-\gamma}{2}} \phi_{s}(k)$. The anomalous dimension
$\gamma$ vanishes in commutative GL model. In the present case,
its finite values is a consequence of the noncommutative effect
\cite{Wu}. Note that only the coupling constant $g_{2}$
contributes to the anomalous dimension.

The anomalous dimension $\gamma$ contains the quartic coupling
constant $u_{2}$, so we should also investigate the RG flow of the
quartic interaction term to one-loop order. The one-loop
correction to the bare quartic term is
\begin{equation}
\frac{s}{8\pi^{2}\Lambda}(1-\frac{1}{s})
\int_{|k^{\prime}|<\Lambda} \phi^{\prime \ast}(1^{\prime})
\phi^{\prime}(2^{\prime}) \phi^{\prime \ast}(3^{\prime})
\phi^{\prime}(4^{\prime}) \mathcal{P}
(1^{\prime}2^{\prime}3^{\prime}4^{\prime})
\end{equation}
with the vertex function $\mathcal{P}
(1^{\prime}2^{\prime}3^{\prime}4^{\prime})$
\begin{eqnarray}
(g_{1}+g_{2})^{2}&(&\frac{1}{2} \cos
(\frac{k^{\prime}_{1}\wedge k^{\prime}_{3}}{2})\cos(\frac{k^{\prime}_{2}\wedge k^{\prime}_{4}}{2}) \nonumber \\
&& + \cos(\frac{k^{\prime}_{1}\wedge k^{\prime}_{2}}{2})\cos(\frac{k^{\prime}_{3}\wedge k^{\prime}_{4}}{2}) \nonumber \\
&&+ \cos(\frac{k^{\prime}_{1}\wedge
k^{\prime}_{4}}{2})\cos(\frac{k^{\prime}_{3}\wedge
k^{\prime}_{2}}{2})).
\end{eqnarray}
Using the identity
\begin{equation}
k^{\prime}_{1}\wedge k^{\prime}_{4} + k^{\prime}_{3}\wedge
k^{\prime}_{2} = - (k^{\prime}_{1}\wedge k^{\prime}_{2} +
k^{\prime}_{3}\wedge k^{\prime}_{4}),
\end{equation}
the vertex function $\mathcal{P}
(1^{\prime}2^{\prime}3^{\prime}4^{\prime})$ reduces to
\begin{eqnarray}
(g_{1}+g_{2})^{2}&(&\frac{3}{2}\cos(\frac{k^{\prime}_{1}\wedge
k^{\prime}_{3}}{2}) \cos(\frac{k^{\prime}_{2}\wedge k^{\prime}_{4}}{2}) \nonumber \\
&& + \cos(\frac{k^{\prime}_{1}\wedge
k^{\prime}_{2}}{2}+\frac{k^{\prime}_{3}\wedge
k^{\prime}_{4}}{2})).
\end{eqnarray}
It has the same form as
$u(1^{\prime}2^{\prime}3^{\prime}4^{\prime})$ in (13), then we can
write down the general coupling parameters for the quartic term as
\begin{eqnarray}
u_{1}^{\prime} & = & su_{1} - \frac{1}{2\pi^{2}} (s^{2}-s)
(u_{1}+u_{2})^{2} \\
u_{2}^{\prime} & = & su_{2} - \frac{3}{4\pi^{2}} (s^{2}-s)
(u_{1}+u_{2})^{2},
\end{eqnarray}
which then leads to the following RG flow equations
\begin{eqnarray}
\frac{du_{1}}{dt} & = & u_{1} - \frac{1}{2 \pi^{2}}(u_{1}+u_{2})^{2} \\
\frac{du_{2}}{dt} & = & u_{2} -
\frac{3}{4\pi^{2}}(u_{1}+u_{2})^{2},
\end{eqnarray}
where $u_{1} = g_{1}\Lambda^{-1}$. It is easy to show that the
fixed point locates at $u_{1}^{\ast} = 8\pi^{2}/25$ and
$u_{2}^{\ast} = 12\pi^{2}/25$.

If the two-point correlation function in the real space $\langle
\phi^{\ast}(\bf x) \phi (0) \rangle \sim {\bf x}^{-1-\gamma}$
tends to zero at large distances, then the fixed point is stable
in the sense that a uniform ordered phase will be formed at zero
temperature. When the noncommutative effect is strong enough, the
correlation function diverges at large distances and a nonuniform
stripe state is reached \cite{Wu}. The critical point that
separates the uniform and non-uniform phases is determined by
$\gamma = -1$.

The anomalous dimension $\gamma$ depends on $\theta \Lambda^{2}$,
so the next step is to choose an appropriate ultraviolet cutoff
$\Lambda$. The noncommutative parameter $\theta$ is determined by
the magnetic length, $\theta = \ell^{2} = \hbar c/eB$. A naive
expectation is to choose the inverse magnetic length as the
ultraviolet cutoff, $\Lambda = \ell^{-1}$, then $\theta
\Lambda^{2} = 1$. We believe that this is not an appropriate
choice since it erases the difference between different LLs.
Instead, we assume that there is an $average$ length scale
$\ell_{e}$ of fermions on {\em all} LLs and define $\Lambda =
\ell_{e}^{-1}$. The effective area each fermion occupies is then
given by $2\pi \ell_{e}^{2}$. The wave pockets of fermions do not
overlap with each other due to the Coulomb repulsion between them.
The total area of the two-dimensional plane is denoted as $A$,
then the actual fermion number is $A/2\pi \ell_{e}^{2}$. The
degeneracy of each LL is $A/2\pi \ell^{2}$. The filling factor is
the ratio of the actual fermion number and the level degeneracy,
thus at $\nu = n/2$ we have
\begin{equation}
\frac{n}{2} = \frac{A/2\pi \ell_{e}^{2}}{A/2\pi\ell^{2}} =
\frac{\ell^{2}}{\ell_{e}^{2}} = \theta \Lambda^{2}.
\end{equation}
Now the anomalous dimension becomes
\begin{equation}
\gamma = -\frac{u_{2}^{\ast}}{6\pi^{2}}(\theta \Lambda^{2})^{2} =
-\frac{n^{2}}{50}.
\end{equation}
It is a remarkable result that the filling factor appears in the
anomalous dimension. Setting $\gamma = -1$ gives the critical
value $n_{c} \simeq 7.07$. Only for $n > n_{c}$, the
noncommutative effect is able to form stripes. This result shows
that stripes exist at $\nu = 9/2, 11/2, 13/2, etc.$ but not at
$\nu = 5/2$ and $7/2$. This is well consistent with transport
experimental observations.

The transport measurements \cite{Lilly, Du99, Fogler01} observed
an anisotropy of resistance at half-fillings of $N \geq 2$ LLs at
temperatures, $T < 0.1$ K. At $\nu = 9/2$, faithful Hall-bar
measurements \cite{Fogler01} found that the anisotropy of
resistances along the two principle directions is as high as
$7:1$. This can be explained as follows. Once the noncommutative
effect drives the formation of stripes along some direction, the
resistance in this direction reduces rapidly down to nearly
vanishing and this kind of stripe is in fact a one-dimensional
superconductor on a two-dimensional noncommutative plane. However,
the resistance in the orthogonal direction is enhanced
significantly because of the small quantum tunnelling between
neighboring stripes. We believe that the noncommutative effect
should play an essential role in forming stripes primarily due to
the experimental fact: when the resistance in one direction
reaches a high peak, the resistance in the orthogonal direction
reaches a deep minimum \cite{Lilly, Du99}. Obviously, there is a
competition between the mobility of carriers in the easy direction
and the hard direction. The space noncommutativity naturally
provides such a competition since the Heisenberg uncertainty
principle requires that the particle is more localized in one
direction, it is more extended in the orthogonal direction.

The transport anisotropy is previously interpreted as the
formation of charge density wave (CDW) state, which was predicted
\cite{Fogler96, Chalker} to exist at half-fillings of high LLs
before the experimental observations. This line of thought tackles
the problem from a microscopic calculation based on the
Hartree-Fock approximation, while we study the phase transition
using the standard RG analysis in an effective mean-field theory.
Within the CDW theory, the ground states of half-fillings of $N =
1$ and $N \geq 2$ LLs are fundamentally different. However, the
experimental fact that an in-plane magnetic field can turn the
$\nu = 5/2$ fractional quantum Hall state into a highly
anisotropic state \cite{Pan} strongly indicates a common property
shared by the ground state of $\nu = 5/2$ and that of $\nu = 9/2,
11/2, 13/2,$ etc.. According to our scenario, they are actually
intimately related with each other. They both undergo a Cooper
pairing instability of composite fermions. The crucial difference
is that the system condenses into a uniform ordering phase at $\nu
= 5/2$ and $7/2$ but is driven by the noncommutative effect to
condense into a nonuniform ordering phase at $\nu = 9/2, 11/2,
13/2,$ etc.. Thus we see that the composite fermion concept offers
a unified description of not only the odd-denominator fractional
quantum Hall effects \cite{Jain, Heinonen} but the half-fillings
of all LLs.

We give a brief discussion of several relevant problems. First,
the RG analysis does not tell us the precise orientation of the
stripe. The orientation should depend on the crystal structure of
materials \cite{Lilly, Du99}. Since the system has a translational
invariance along the stripe direction, the stripe would prefer to
align itself in the direction that respects this symmetry. Second,
the anisotropy of resistance was found \cite{Lilly, Du99} to be
largest at $\nu = 9/2$ and decrease with growing filling factor
index $n$. The reason is presumably that as $n$ grows, the quantum
fluctuation increases, resulting in increasing quantum tunnelling
between stripes. In addition, it was found that a strong in-plane
magnetic field can pin down the orientation of stripes \cite{Pan}.
It is currently not known how to understand this experimental fact
using the noncommutative effect.

It is interesting to notice a similarity of the phase transition
to stripe state driven by the uncertainty relation of spatial
coordinates to the phase transition from a Mott insulator with the
local particle number being fixed to a superfluid with the phase
of boson wave function being fixed \cite{Greiner}. The latter is
controlled by the Heisenberg uncertainty relation between the
local particle number and the phase of the boson wave function.
The stripe state, Mott insulator state and superfluid state all
correspond to a limit that one physical variable takes a fixed
value while the conjugate variable that does not commute with it
is completely uncertain.

The space noncommutativity is a profound concept in physics. Only
experiments can tell us whether the real world is a noncommutative
space or not. We hope the results in this paper shed some light on
the search for space noncommutativity in nature.

G.Z.L would like to thank Yong-Shi Wu, Zheng-Wei Wu and Fei Xu for
very helpful discussions. G.Z.L. is supported by the NSF of China
No. 10404024.


\begin{thebibliography}{99}

\bibitem{Jain}
J. K. Jain, Phys. Rev. Lett. {\bf 62}, 199 (1989).

\bibitem{Heinonen}
O. Heinonen (ed.), {\em Composite Fermion} (World Scientific, New
York, 1998).

\bibitem{Halperin}
B. I. Halperin, P. A. Lee, and N. Read, Phys. Rev. B {\bf 47},
7312 (1993).

\bibitem{Willett1}
R. Willett $et$ $al.$, Phys. Rev. Lett. {\bf 59}, 3846 (1993).

\bibitem{Willett2}
R. Willett $et$ $al.$, Phys. Rev. Lett. {\bf 59}, 1776 (1987).

\bibitem{Moore}
G. Moore and N. Read, Nucl. Phys. B {\bf 360}, 362 (1991).

\bibitem{Greiter}
M. Greiter, X.-G. Wen and F. Wilczek, Phys. Rev. Lett. {\bf 66},
3205 (1991); Nucl. Phys. B {\bf 374}, 567 (1992).

\bibitem{Morf}
R. Morf, Phys. Rev. Lett. {\bf 80}, 1505 (1998); E. H. Rezayi and
F. D. M. Haldane, Phys. Rev. Lett. {\bf 84}, 4685 (2000).

\bibitem{Scarola}
V. W. Scarola, K. Park, and J. K. Jain, Nature (London) {\bf 406},
863 (2000).

\bibitem{Lilly}
M. P. Lilly $et$ $al.$, Phys. Rev. Lett. {\bf 82}, 394 (1999).

\bibitem{Du99}
R. R. Du $et$ $al.$, Solid State Commun. {\bf 109}, 389 (1999).

\bibitem{Girvin}
S. M. Girvin and T. Jach, Phys. Rev. B {\bf 29}, 5617 (1984).

\bibitem{Bellissard}
J. Bellissard, A. van Elst, and H. Schultz-Baldes,
cond-mat/9411052.

\bibitem{Snyder}
H. S. Snyder, Phys. Rev. {\bf 71}, 38 (1947).

\bibitem{Douglas}
M. R. Douglas and N. A. Nekrasov, Rev. Mod. Phys. {\bf 73}, 977
(2001).

\bibitem{Connes94}
A. Connes, {\em Noncommutative Geometry} (Academic Press, Inc.;
New York, 1994).

\bibitem{Sondhi}
S. S. Gubser and S. L. Sondhi, Nucl. Phys. B {\bf 605}, 395
(2001).

\bibitem{Wu}
G.-H. Chen and Y.-S. Wu, Nucl. Phys. B {\bf 622}, 189 (2002).

\bibitem{Shankar}
R. Shankar, Rev. Mod. Phys. {\bf 66}, 129 (1994).

\bibitem{Polchinski}
J. Polchinski, in {\em Proceedings of 1992 Theoretical Advanced
Studies Institute in Elementary Particle Physics}, edited by J.
Harvey and J. Polchinski (World Scientific, Singapore, 1993).

\bibitem{Luban}
R. M. Hornreich, M. Luban, and S. Shtrikman, Phys. Rev. Lett. {\bf
35}, 1678 (1975).



\bibitem{Fogler01}
For a review, see M. M. Fogler, cond-mat/0111001.

\bibitem{Fogler96}
A. A. Koulakov, M. M. Fogler, and B. I. Shklovskii, Phys. Rev.
Lett. {\bf 76}, 499 (1996); M. M. Fogler, A. A. Koulakov, and B.
I. Shklovskii, Phys. Rev. B {\bf 54}, 1853 (1996).

\bibitem{Chalker}
R. Moessner and J. T. Chalker, Phys. Rev. B {\bf 54}, 5006 (1996).

\bibitem{Pan}
W. Pan $et$ $al.$, Phys. Rev. Lett. {\bf 83}, 820 (1999); M. P.
Lilly $et$ $al.$, Phys. Rev. Lett. {\bf 83}, 824 (1999).

\bibitem{Greiner}
M. Greiner $et$ $al.$, Nature (London) {\bf 415}, 39 (2002).

\end{thebibliography}
\end{document}